\documentclass{emulateapj}

\begin{document}
\slugcomment{Accepted by ApJ}

\title{ Hypervelocity Stars. I. The Spectroscopic Survey}

\author{Warren R.\ Brown\altaffilmark{1},
	Margaret J.\ Geller,
	Scott J.\ Kenyon, and
	Michael J.\ Kurtz}

\affil{Smithsonian Astrophysical Observatory, 60 Garden St, Cambridge, MA 02138}
\email{wbrown@cfa.harvard.edu}

\altaffiltext{1}{Clay Fellow, Harvard-Smithsonian Center for Astrophysics}

\shorttitle{ Hypervelocity Stars. I. }
\shortauthors{Brown et al.}

\begin{abstract}

	We discuss our targeted search for hypervelocity stars (HVSs), stars
traveling with velocities so extreme that dynamical ejection from a massive
black hole is their only suggested origin.  Our survey, now half complete, has
successfully identified a total of four probable HVSs plus a number of other
unusual objects.  Here we report the most recently discovered two HVSs:  SDSS
J110557.45+093439.5 and possibly SDSS J113312.12+010824, traveling with
Galactic rest-frame velocities at least $+508\pm12$ and $+418\pm10$ km
s$^{-1}$, respectively.  
	The other late B-type objects in our survey are consistent with a
population of post-main sequence stars or blue stragglers in the Galactic
halo, with mean metallicity [Fe/H]$_{W_k}=-1.3$ and velocity dispersion
$108\pm5$ km s$^{-1}$.  Interestingly, the velocity distribution shows a tail
of objects with large positive velocities that may be a mix of low-velocity
HVSs and high-velocity runaway stars.  Our survey also includes a number of DA
white dwarfs with unusually red colors, possibly extremely low mass objects.  
Two of our objects are B supergiants in the Leo A dwarf, providing the first
spectroscopic evidence for star formation in this dwarf galaxy within the last
$\sim$30 Myr.

\end{abstract}


\section{INTRODUCTION}

	Hypervelocity stars (HVSs) travel with velocities so extreme that
dynamical ejection from a massive black hole (MBH) is their only suggested
origin.  First predicted by \citet{hills88}, HVSs traveling $\sim$1,000 km
s$^{-1}$ are a natural consequence of a MBH in a dense stellar environment like
that in the Galactic center.  HVSs differ from runaway stars because 1) HVSs
are unbound and 2) the classical supernova ejection \citep{blaauw61} and
dynamical ejection \citep{poveda67} mechanisms that explain runaway stars
cannot produce ejection velocities larger than 200 - 300 km s$^{-1}$
\citep{leonard91,leonard93,portegies00,gualandris04,dray05}.  Depending on the
actual velocity distributions of HVSs and runaway stars, some HVSs ejected by
the central MBH may overlap with runaway stars in radial velocity.

	Following the original prediction of HVSs, \citet{hills91} provided a
comprehensive analysis of orbital parameters needed to produce HVSs, and
\citet{yu03} expanded the \citet{hills88} analysis to include the case of a
binary black hole and to predict HVS production rates.  In 2005, Brown and
collaborators reported the first discovery of a HVS:  a $g'=19.8$ B9 star,
$\sim$110 kpc distant in the Galactic halo, traveling with a Galactic
rest-frame velocity of at least $+709\pm12$ km s$^{-1}$ (heliocentric radial
velocity +853 km s$^{-1}$).  Photometric follow-up revealed that the object is
a slowly pulsating B main sequence star \citep{fuentes06}.  Only interaction
with a MBH can plausibly accelerate a 3 $M_{\sun}$ main sequence B star to such
an extreme velocity.

	The discovery of the first HVS inspired a wealth of theoretical and
observational work.  Because HVSs originate from a close encounter with a MBH,
HVSs can be used as important tools for understanding the nature and environs
of MBHs \citep{gualandris05,levin05,ginsburg06,holley06,demarque06}.  The
trajectories of HVSs also provide unique probes of the shape and orientation of
the Galaxy's dark matter halo \citep{gnedin05}.  Recent discoveries of new HVSs
\citep{edelmann05,hirsch05,brown06} are starting to allow observers to place
suggestive limits on the stellar mass function of HVSs, the origin of massive
stars in the Galactic Center, and the history of stellar interactions with the
MBH.  Clearly, a larger sample of HVSs will be a rich source for further
progress on these issues.

	Here we discuss our targeted survey for HVSs and the unusual objects we
find in it.  To discover HVSs, we have undertaken a radial velocity survey of
faint B-type stars, stars with lifetimes consistent with travel times from the
Galactic center but which are not a normally expected stellar halo population.  
This strategy is successful:  approximately 1-in-50 of our candidate B stars
are HVSs.  The first two HVS discoveries from our survey are presented in
\citet{brown06}.  Here we present two further HVS discoveries -- one certain
HVS and one possible HVS.  In addition to HVSs, our survey has uncovered many
unusual objects with late B-type colors:  post-main sequence stars, young B
supergiant stars, DA, DB, and DZ white dwarfs, and one extreme low-metallicity
starburst galaxy.

	Our paper is organized as follows.  In \S 2 we discuss the target
selection and spectroscopic identifications of objects in our survey, now half
complete.  In \S 3 we present two new HVS discoveries.  In \S 4 we show that
the properties of the other late-B type stars in the sample are consistent with
being a Galactic halo population of post-main sequence stars and/or blue
stragglers.  In \S 5 we discuss the white dwarfs in the sample, many of which
may be unusually low mass DA white dwarfs.  In \S 6 we discuss two young B
supergiants in the Leo A dwarf galaxy and one UV bright phase horizontal branch
star in the Draco dwarf galaxy.  We conclude in \S 7.

\section{DATA}

\subsection{Target Selection}

	As \citet{brown06} discuss, HVSs ought to be rare:  \citet{yu03}
predict there should be $\sim$10$^3$ HVSs in the entire Galaxy.  Thus, in any
search for HVSs, {\it survey volume is important}.  Solar neighborhood surveys
have not discovered HVSs because, even if they were perfectly complete to a
depth of 1 kpc, there is only a $\sim$0.1\% chance of finding a HVS in such a
small volume.  Finding a new HVS among the Galaxy's $\sim$10$^{11}$ stars also
requires selection of targets with a high probability of being HVSs.  Our
observational strategy is two-fold.  Because the density of stars in the
Galactic halo drops off as approximately $r^{-3}$, and the density of HVSs
drops off as $r^{-2}$ (if they are produced at a constant rate), we target
distant stars where we {\it maximize the contrast} between the density of HVSs
and indigenous stars.  Secondly, the stellar halo contains mostly old,
late-type stars.  Thus we target faint B-type stars, stars with lifetimes
consistent with travel times from the Galactic center but which are not a
normally expected stellar halo population.  O-type stars are more luminous but
do not live long enough to reach the halo.  A-type stars are also luminous but
must be detected against large numbers of evolved blue horizontal branch (BHB)  
stars in the halo.  Based on the \citet{brown05b} field BHB luminosity
function, we expect only small numbers of hot BHB stars with B-type colors.  
Our strategy of targeting B-type stars is further supported by observations
showing that 90\% of the $K<16$ stars in the central $0.5\arcsec$ of the
Galactic Center are in fact normal main sequence B stars \citep{eisenhauer05}.

\begin{figure}          
 \includegraphics[width=3.25in]{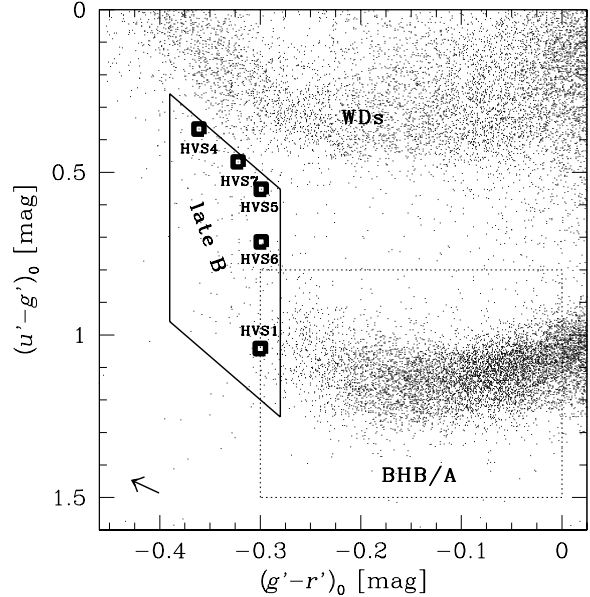}
 \figcaption{ \label{fig:ugr}
        Color-color diagram showing our target selection, illustrated with   
every star in the SDSS DR4 $17.5<g'_0<18.5$.  For reference, BHB/A stars are
located in the dotted box \citep{yanny00}.  Candidate B-type stars extend up   
the stellar sequence towards the ensemble of white dwarfs, and are selected
within the solid parallelogram.  The arrow indicates the amplitude and
direction of the median reddening correction for our targets.  Open squares  
mark the HVSs we have discovered \citep{brown05,brown06} and the two new HVSs
presented here.} 
\end{figure}

	We use Sloan Digital Sky Survey (SDSS) photometry to select candidate B
stars by color.  Our color selection is illustrated in Fig.\ \ref{fig:ugr}, a
color-color diagram of stars with B- and A-type colors in the SDSS Fourth Data
Release \citep{adelman06}.  \citet{fukugita96} describe the SDSS filter system
and the colors of main sequence stars in the SDSS photometric system.  We use
SDSS point-spread function magnitudes and reject any objects that have bad
photometry flags.  We compute de-reddened colors using extinction values
obtained from \citet{schlegel98}.  The dotted box in Fig.\ \ref{fig:ugr}
indicates the selection region used by \citet{yanny00} to identify BHB
candidates.  Interestingly, there is a faint group of stars with late B-type
colors extending up the stellar sequence towards the ensemble of white dwarfs.  
We chose our primary candidate B star selection region inside the solid
parallelogram defined by: $-0.38<(g'-r')_0<-0.28$ and $2.67(g'-r')_0 + 1.30 <
(u'-g')_0 < 2.67(g'-r')_0 + 2.0$.  In addition, we impose $-0.5<(r'-i')_0<0$ to
reject objects with non-stellar colors.

	We observe candidate B stars in the magnitude range $17.0<g'_0<19.5$.  
The bright magnitude limit sets an inner distance boundary $\gtrsim$30 kpc for
late B-type stars, a distance beyond that of known runaway B stars
\citep{lynn04,martin04}.  We chose the faint magnitude limit to keep our
exposure times below 30 minutes using the 6.5m MMT telescope.  In addition, we
exclude the region of sky between $b<-l/5 + 50\arcdeg$ and $b>l/5-50\arcdeg$ to
avoid excessive contamination from Galactic bulge stars.

\begin{figure*}          
 \includegraphics[width=7.0in]{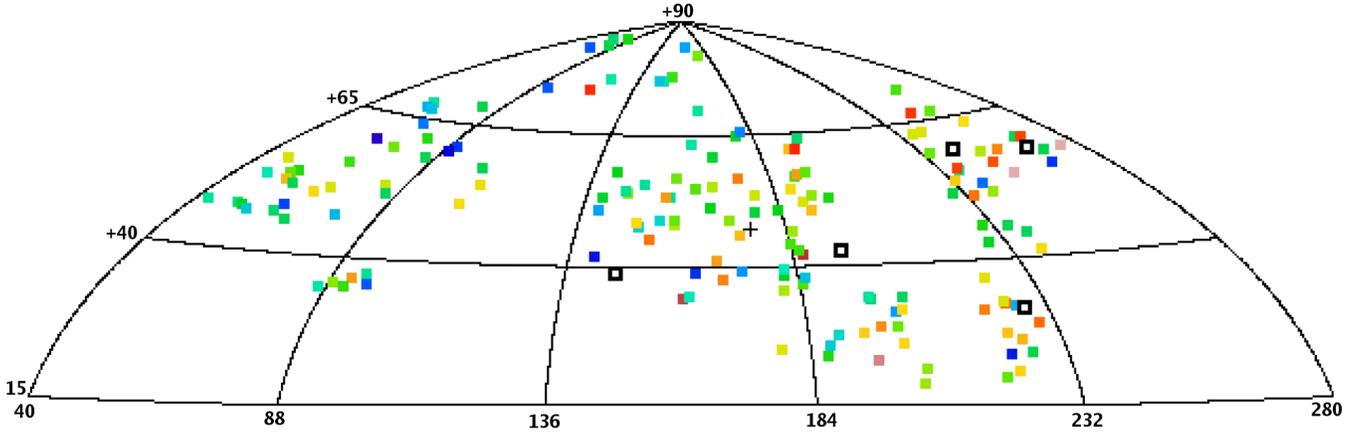}
 \figcaption{ \label{fig:sky}
        Aitoff sky map, in Galactic coordinates, showing the observed candidate
B stars.  Radial velocities, in the Galactic rest frame, are indicated by the 
color of the solid squares:  purple is -300, green is 0, and red is +300 km
s$^{-1}$.  Our HVSs (see Table \ref{tab:hvs}) are completely off this color   
scale and are marked by open squares; HVS2 \citep{hirsch05} is marked by a plus
sign. }
 \end{figure*}

	There are a total of 430 SDSS DR4 candidate B stars in the primary
selection region described above.  We have observed 192, or 45\% of this total.  
The average surface number density of targets is 1 per 15 deg$^2$.  Thus we
have surveyed $\sim$3000 deg$^2$ or 7\% of the entire sky.  Figure
\ref{fig:sky} displays the locations of observed candidate B stars in the
northern Galactic hemisphere;  a handful of stars in the autumn SDSS equatorial
stripes are located in the southern Galactic hemisphere and are not displayed.

	In addition, we have observed 55 targets with colors, magnitudes, or
positions slightly outside our primary selection region (for example, see Fig.\
\ref{fig:ugr2}).  We include the full sample of 247 objects in our discussion
below.  We note that the region of sky located $40<l<90^{\circ}$ in Fig.\
\ref{fig:sky} lacks HVS discoveries, but this is probably not significant.
This region, observed in July 2005, is missing observations of the bluest 
objects in $(u'-g')_0$ and will be completed in the coming months.

\subsection{Spectroscopic Observations and Radial Velocities}

	Observations were obtained with the Blue Channel spectrograph on the
6.5m MMT telescope.  Observations were obtained on the nights of 2005 July
10-11, 2005 December 3-5, and 2006 February 22-25.  The spectrograph was
operated with the 832 line/mm grating in second order, providing wavelength
coverage 3650 \AA\ to 4500 \AA.  Most spectra were obtained with 1.2 \AA\
spectral resolution, however on one night of poor seeing we used a larger slit
that provided 1.5 \AA\ spectral resolution for 24 objects.  Exposure times
ranged from 5 to 30 minutes and were chosen to yield $S/N=15$ in the continuum
at 4000 \AA.  Comparison lamp exposures were after obtained after every
exposure.

	Radial velocities were measured using the cross-correlation package
RVSAO \citep{kurtz98}.  \citet{brown03} describe in detail the
cross-correlation templates we use.  Errors are measured from the width of the
cross-correlation peak, and are added in quadrature with the 9 km s$^{-1}$
systematic uncertainty observed in bright BHB standards.  The average
uncertainty is $\pm11$ km s$^{-1}$ for the late B-type stars and $\pm40$ km
s$^{-1}$ for the DA white dwarfs (with much broader Balmer lines).

\subsection{Selection Efficiency and Unusual Objects}

	Our candidate B stars include post-main sequence stars and late B blue
stragglers, some DA white dwarfs, and a few other unusual objects.  We classify
the spectral types of the 202 late B stars based on \citet{oconnell73} and
\citet{worthey94} line indices as described in \citet{brown03}.  The spectral
types of the stars range from B6 to A1 with an average uncertainty of $\pm1.6$
spectral sub-types.  Thus our primary target selection is 84\% efficient for
selecting stars of late B spectral type.  Four of the 202 late B stars, or
approximately 1-in-50, are HVSs.  In addition, 3 of the late B stars coincide
with Local Group dwarf galaxies, which provides special constraints on the
nature of those objects.

\begin{figure}          
 \includegraphics[width=3.25in]{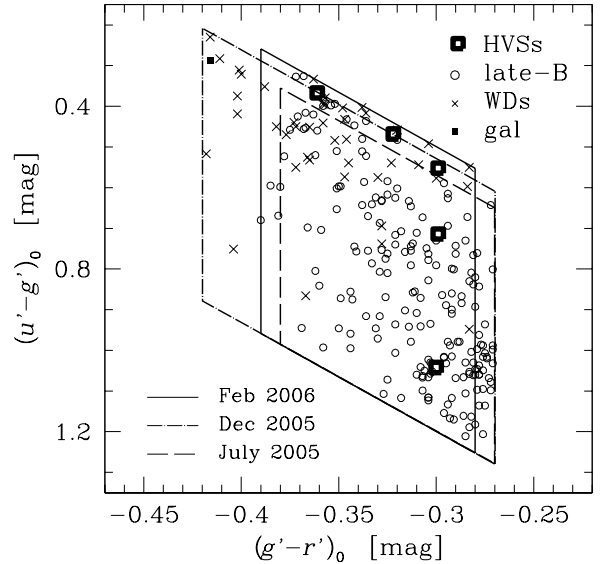}
 \figcaption{ \label{fig:ugr2}
        Color-color diagram showing spectroscopic identifications for the full
sample of objects we observed with the MMT.  The primary color-selection region
is indicated by the solid line; dashed lines show the selection regions used
during the December and July observing runs.  The primary selection region is
84\% efficient for selecting late B-type stars, of which approximately 1-in-50
are HVSs.}
 \end{figure}

	Figure \ref{fig:ugr2} plots the colors and spectroscopic
identifications for the full sample of objects.  The solid parallelogram
indicates our primary color selection region; the dashed lines show the
slightly different color selection regions used on different observing runs.  
44 of the objects in Fig.\ \ref{fig:ugr2} are white dwarfs marked by crosses.  
The white dwarfs are mostly DA white dwarfs, but also include one DB and one DZ
white dwarf.  Our sample also includes one extreme low-metallicity starburst
galaxy, marked by the solid square in Fig.\ \ref{fig:ugr2}, that we describe in 
a separate paper (Kewley et al., in preparation).

\section{HYPERVELOCITY STARS}

	Our targeted search for HVSs has discovered a total of four probable
HVSs.  \citet{brown06} report the discovery of the first two HVSs from this
survey, and here we report two further HVS discoveries:  SDSS
J110557.45+093439.5 (hereafter HVS6) and possibly SDSS J113312.12+010824.9
(hereafter HVS7).  HVS6 is a faint $g'=19.06\pm0.02$ star with B9 spectral type
and travels with a $+606\pm12$ km s$^{-1}$ heliocentric radial velocity.  HVS7
is a $g'=17.75\pm0.02$ star with B7 spectral type and travels with a
$+531\pm10$ km s$^{-1}$ heliocentric radial velocity.  We correct the
velocities to the local standard of rest \citep{hogg05} and remove the 220 km
s$^{-1}$ solar reflex motion as follows:
	\begin{equation}
 v_{rf} = v_{helio} + (10\cos{l}\cos{b} + 5.2\sin{l}\cos{b} + 7.2\sin{b}) + 220\sin{l}\cos{b}
	\end{equation} The minimum Galactic rest frame velocities (indicated
$v_{rf}$) of HVS6 and HVS7 are $+508$ and $+418$ km s$^{-1}$, respectively.  
The minimum Galactic rest-frame velocity of HVS7 is marginally consistent with
runaway star mechanisms, but, if it is a main sequence B star, it is unbound
to the Galaxy.  Thus, for now, we consider HVS7 a HVS.  All 7 known HVSs are
traveling with large positive radial velocity, consistent with a Galactic
center origin.

\begin{figure}          
 \includegraphics[width=3.25in]{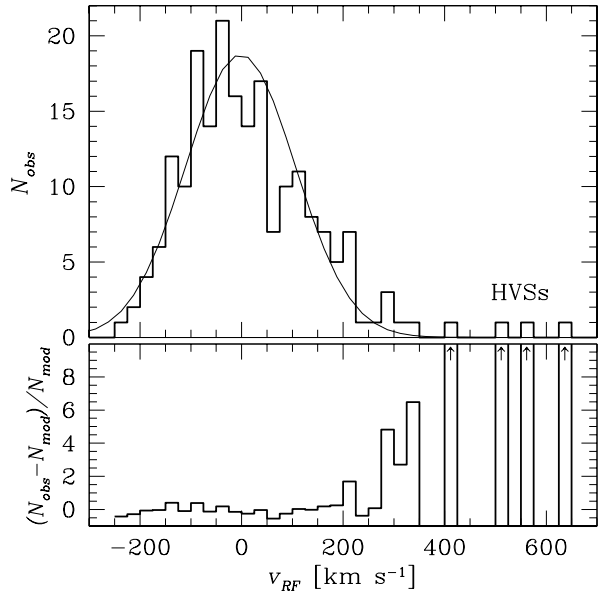}
 \figcaption{ \label{fig:velh}
        Galactic rest-frame velocity histogram of the late B-type stars ({\it
upper panel}).  The best-fit Gaussian ({\it thin line}) has dispersion
$108\pm5$ km s$^{-1}$.  Our survey has identified a total of four HVSs that are
4-6$\sigma$ outliers from this distribution.  The lower panel plots the   
residuals of the observations from the best-fit Gaussian, normalized by the
value of the Gaussian.  In addition to the HVSs, there is an interesting tail  
of high positive velocity objects $200<v_{rf}<400$ km s$^{-1}$.}
 \end{figure}

	Figure \ref{fig:velh} plots a histogram of Galactic rest-frame velocity
for the 202 late B stars in our sample.  We calculate the line-of-sight
velocity dispersion of the stars using three different methods:  1) fitting a
Gaussian to the entire distribution, 2) fitting a Gaussian to just the negative
velocity half of the distribution, and 3) simply calculating the dispersion
around the mean after clipping the HVSs.  All three methods yield equivalent
results.  Averaging the results of the three methods, our sample has a velocity
dispersion of $108\pm5$ km s$^{-1}$ and mean of $-2\pm8$ km s$^{-1}$ consistent
with a Galactic halo population.

	HVS6 and HVS7 are 4.7$\sigma$ and 3.9$\sigma$ outliers, respectively,
from the velocity distribution.  The lower panel of Fig.\ \ref{fig:velh} plots
the residuals of the observations from the best-fit Gaussian, normalized by the
value of the Gaussian.  Stars with velocities below $|v_{rf}|<200$ km s$^{-1}$
show low-significance deviations from a Gaussian distribution.  The four HVSs,
on the other hand, are 4-6$\sigma$ outliers and are completely off-scale.

	In addition to the HVSs, the distribution of velocities in Fig.\
\ref{fig:velh} shows a tail of stars traveling with large positive velocities
$v_{rf}>250$ km s$^{-1}$ and no stars traveling with equally large negative
velocities.  Stars in compact binary systems may produce outliers in the
velocity distribution, but such outliers should be distributed symmetrically.  
Conceivably, the observed asymmetry is the low-velocity tail of HVSs, or it may
be the high-velocity tail of runaway stars.  Because runaway stars are ejected
with low $<300$ km s$^{-1}$ velocities, they follow bound, ballistic
trajectories away from and then back onto the disk \citep[e.g.][]{martin06}.  
Thus, if the stars in the high velocity tail are runaway stars, they must be
very nearby.  The exact velocity distribution of runaway stars is, however,
unclear.  The predictions of \citet{portegies00} are not applicable, for
example, because the runaway stars must be low mass, intrinsically faint
objects to be located nearby. Moreover, the velocity distribution of HVSs has
been calculated for only restrictive sets of circumstances
\citep{hills91,levin05,ginsburg06}.  Clean predictions of runaway star and HVS
velocity distributions are needed to discriminate among the populations in the
high velocity tail.  Proper motions (as may be measured with the {\it Hubble
Space Telescope}, {\it GAIA}, or the {\it Space Interferometry Mission}) will
ultimately discriminate between HVSs and runaway stars.

	Our low-resolution spectra do not allow determination of exact stellar
parameters for HVS6 and HVS7.  Stars of late B spectral type are probably
post-main sequence stars or main sequence B stars/blue stragglers.  We note
that the Balmer line widths of HVS6 and HVS7 are too broad to be consistent
with those of luminosity class I or II B supergiants.  If we assume the HVSs
are BHB stars rather than B stars, their blue colors mean they are hot, extreme
BHB stars and thus they are intrinsically very faint. The $M_V(BHB)$ relation
of \citet{clewley02} yields $M_V(BHB)\simeq+1.6$ and $+1.8$ and heliocentric
distance estimates $d_{BHB}\simeq$30 and 15 kpc for HVS6 and HVS7,
respectively.  In the BHB interpretation, the volume we effectively survey is
much smaller than in the B star interpretation.  Because the first two HVSs are
known B stars \citep{edelmann05,fuentes06} and because the B star
interpretation implies a production rate probably consistent with \citet{yu03},
we assume that HVS6 and HVS7 are B stars for the purpose of discussion.  The
ultimate discriminant will come from higher resolution, higher signal-to-noise
spectroscopy.

	We estimate distances for HVS6 and HVS7 by looking at
\citet{schaller92} stellar evolution tracks for 3 and 4 $M_{\sun}$ stars with
$Z=0.02$.  A 3 $M_{\sun}$ star spends 350 Myr on the main sequence with $M_V(3
M_{\sun})\simeq-0.3$.  If HVS6 is a 3 $M_{\sun}$ B9 main sequence star, it has
a heliocentric distance $d\sim75$ kpc.  Using this distance, we now
estimate the HVS travel time from the Galactic Center.  We make the
conservative assumptions that the HVS's observed velocity is a total space
velocity and that its velocity has remained constant.  Detailed calculations of
HVS trajectories by \citet{gnedin05} show that this simple estimate is
reasonably accurate and over-estimates HVS travel times by less than 10\% (O.\
Gnedin, private communication).  We estimate the travel time of HVS6 is
$\sim$160 Myr, consistent with its 350 Myr main sequence lifetime.  By
comparison, a 4 $M_{\sun}$ star spends 160 Myr on the main sequence and has
$M_V(4 M_{\sun})\simeq-0.9$.  If HVS7 is a 4 $M_{\sun}$ B7 main sequence star,
it has a heliocentric distance $d\sim55$ kpc and a travel time from the
Galactic center of $\sim$120 Myr also consistent with its lifetime.  There is a
tendency to find HVSs near the end of their lives because the longer they have
traveled, the larger the survey volume they populate and the greater the
contrast with the indigenous stellar populations.

	Our radial velocities provide only a {\it lower} limit to the HVSs'
true space velocities.  The escape velocity from the Galaxy is approximately
300 km s$^{-1}$ at 50 kpc \citep{wilkinson99}, thus HVS6 is unbound to the
Galaxy whether it is a B main sequence star or a BHB star.  HVS7, on the other
hand, is only unbound if it is a B main sequence star; follow-up spectroscopy
is necessary to establish whether it is a ``true'' HVS.

	HVS6 and HVS7 are both present in the USNOB1 \citep{monet03} catalog
but only HVS7 is listed with a proper motion.  Averaging the USNOB1 proper
motion with that from the GSC2.3 (B.\ McLean, 2006 private communication), HVS7
has $\mu=10.5\pm9$ mas yr$^{-1}$.  If we assume HVS7 is located nearby at
$d_{BHB}\simeq$15 kpc consistent with a proper motion detection, then its
transverse velocity is $750\pm650$ km s$^{-1}$.  Such a velocity would suggest
that HVS7 is unbound, but the proper motion measurement is significant at only
the 1$\sigma$ level and thus we place little confidence in it.

	The new HVSs are not physically associated with any other Local Group
galaxy.  HVS6 and HVS7 are located at $(l,b)=(243\fdg1,59.\fdg6)$ and
$(263\fdg8,57\fdg9)$, respectively (see Fig.\ \ref{fig:sky}).  The nearest
galaxies to HVS6 are Leo I and Leo II, both $\sim$14$^{\arcdeg}$ away on the
sky from HVS6.  However, Leo I and Leo II are at distances of $254\pm17$ kpc
\citep{bellazzini04} and $233\pm15$ \citep{bellazzini05}, respectively, many
times the estimated distance of HVS6.  Thus HVS6 is moving towards Leo I and
Leo II at minimum velocities of 330 km s$^{-1}$ and 490 km s$^{-1}$,
respectively, and clearly unrelated to those galaxies.  The nearest galaxy to
HVS7 is the Sextans dwarf 20$^{\arcdeg}$ away on the sky.  At a distance of
$1320\pm40$ kpc \citep{dolphin03}, Sextans is unrelated to HVS7.

	Table \ref{tab:hvs} summarizes the properties of all seven known HVSs,
four of which were discovered in this survey.  The columns include HVS number,
Galactic coordinates $(l,b)$, apparent magnitude $g'$, minimum Galactic
rest-frame velocity $v_{RF}$ (not a full space velocity), heliocentric distance
estimate $d$, travel time estimate from the Galactic Center $t_{GC}$, and
catalog identification.  We have repeat observations of HVS1, HVS4, and HVS5;
their radial velocities are constant within the uncertainties.

\section{HALO STARS}

	Most objects in our survey are halo stars with late B spectral types,
and we now discuss their nature.  Stars of late B spectral type are probably
main sequence stars / blue stragglers or post-main sequence stars.  
Unfortunately, main sequence stars and post-main sequence BHB stars share
similar effective temperature (color) and surface gravity (spectral line
widths), making classification difficult.

	Stellar rotation is a useful discriminant between rapidly rotating main
sequence B stars \citep{abt02,martin04} and slowly rotating BHB stars
\citep{peterson95,behr03}.  However, our low-dispersion spectra do not allow us
to measure rotation.  Instead, we constrain the nature of the late B type
objects by looking at their metallicities and kinematics.

\begin{figure}          
 \includegraphics[width=3.25in]{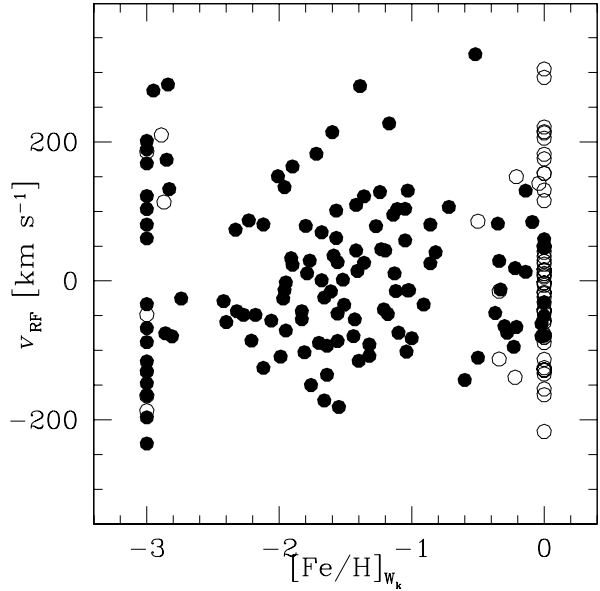}
 \figcaption{ \label{fig:velfeh}
        Metallicities and Galactic rest frame velocities of the late B stars.   
The \citet{wilhelm99a} models restrict our metallicity estimates to $-3 <$
[Fe/H]$_{W_k}$ $< 0$.  These metallicities assume the stars have $\log{g}=3$;
using $\log{g}=4$ does not substantially change the distribution.  Hot
objects with especially poor [Fe/H]$_{W_k}$ determinations are plotted as open
symbols.} 
\end{figure}

	The strongest indicator of metallicity in our spectra is the 3933 \AA\
Ca{\sc ii} K line.  The equivalent width of Ca{\sc ii} K depends on both
temperature and metallicity.  To estimate metallicity, we first compute
$(\bv)_0$ from the SDSS colors following \citet{clewley05}.  We then measure
the equivalent width of the Ca{\sc ii} K line, $W_k$.  Finally, we estimate
metallicity [Fe/H]$_{W_k}$ by interpolating the theoretical curves of
\citet{wilhelm99a}, assuming $\log{g}=3$ appropriate for a BHB star.  We
propagate the errors in $(\bv)_0$ and $W_k$ through the \citet{wilhelm99a}
curves and find that the uncertainty is large, $\pm0.67$ in [Fe/H]$_{W_k}$.  
Moreover, Ca{\sc ii} K provides very little leverage on metallicity for the
hottest stars $(\bv)_0<-0.05$.  For the 105 stars with $(\bv)_0<-0.05$ our
metallicity estimates are effectively reduced to a binary measurement:  
[Fe/H]$_{W_k}$$\sim$0 if we see Ca{\sc ii} K, and [Fe/H]$_{W_k}$$\sim-3$ if
Ca{\sc ii} K is absent.  Note that the \citet{wilhelm99a} models restrict our
metallicity estimates to $-3 <$ [Fe/H]$_{W_k}$ $< 0$.

	Figure \ref{fig:velfeh} plots metallicities and Galactic rest frame
velocities of the late B type objects.  We plot hot objects with poor
[Fe/H]$_{W_k}$ determinations as open symbols.  All of the objects are affected
at some level by accretion of interstellar material, atomic diffusion in their
atmospheres, and observational effects such as interstellar absorption.  We do
not know the detailed histories of the individual stars, and thus we simply
consider the average observed [Fe/H]$_{W_k}$ of the sample. Ignoring the
objects on the [Fe/H]$_{W_k}=0$ and $-3$ boundaries, it is clear that the
objects cluster at metal-poor values:  the mean metallicity of the sample
(excluding objects on the boundaries) is $\overline{{\rm[Fe/H]}_{W_k}}=-1.3$.  
If instead we assume $\log{g}=4$ appropriate for main sequence/blue straggler
stars, additional stars are pushed onto the [Fe/H]$_{W_k}=0$ boundary line and
the mean metallicity of the sample (excluding objects on the boundaries)
increases slightly to $\overline{{\rm[Fe/H]}_{W_k}}=-1.2$.  The low mean
metallicity of the sample suggests that most objects are not recently-formed
main sequence B stars ejected from the disk, but rather the objects are likely
post-main sequence stars or old blue stragglers.

	The observed $108\pm5$ km s$^{-1}$ velocity dispersion of the late B
type objects is also consistent with a Galactic halo population of post-main
sequence stars or blue stragglers.  Although some have proposed in-situ star
formation in the halo \citep{vanwoerden93,christodoulou97}, there is no
evidence for this in modern studies of runaway B stars, including the recent
\citet{martin06} study of runaway stars in the Hipparcos catalog.  Thus the
observed metallicity and velocity distributions suggest that the late B type
stars are most likely a Galactic halo population of post-main sequence stars
and/or old blue stragglers, and not young runaway B stars ejected from the
disk.  We hope ultimately to use this sample to provide a useful probe of halo
structure.


	Table \ref{tab:bhb} lists the 202 spectroscopically identified late B
type objects, including the 4 HVSs.  The columns include RA and Dec coordinates
(J2000), $g'$ apparent magnitude, $(u'-g')_0$ and $(g'-r')_0$ color, and our
heliocentric velocity $v_{helio}$ and [Fe/H]$_{W_k}$ estimate.

\section{WHITE DWARFS}

	44 survey objects are faint white dwarfs, drawn from a largely
unexplored region of color-space compared to previous SDSS-based white dwarf
spectroscopic surveys \citep{harris03,kleinman04,kilic06}.  The objects are
almost entirely DA white dwarfs, with colors $-0.4<(u'-g')_0<0.2$ indicating
temperatures $10,000 < T_{eff} < 16,000$ K \citep{kleinman04}.  Our color
selection region, however, lies at surface gravities $\log{g}<7$ for
hydrogen-atmosphere white dwarfs (i.e.\ to the right of the Bergeron
$\log{g}=7$ curve plotted in Fig.\ 1 of \citet{harris03}).  Thus the white
dwarfs we find are all candidates for objects with unusually low surface
gravities and unusually low masses.  

\begin{figure}          
 \includegraphics[width=3.25in]{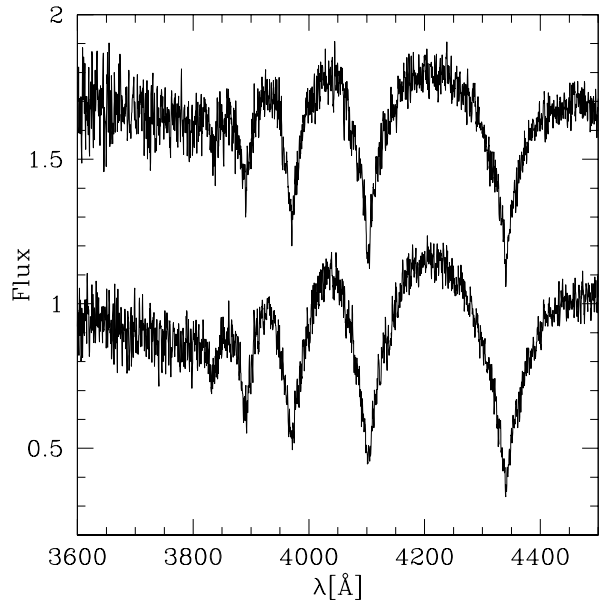}
 \figcaption{ \label{fig:wd}
        Spectra of two DA white dwarfs, SDSS J074508.15+182630.0 ({\it upper})
and SDSS J083303.03+365906.3 ({\it lower}), having unusually red $(u'-g')_0$
colors and possibly unusually low mass.}
 \end{figure}

	The least massive white dwarfs known are $\sim$0.2 $M_{\sun}$
helium-core objects in binary systems containing millisecond pulsars
\citep[e.g.][]{callanan98} or subluminous B (sdB) stars \citep{heber03,
otoole06}.  \citet{liebert04} discuss a 0.18 $M_{\sun}$ helium white dwarf in
the SDSS with colors $(u'-g')=0.32$ and $(g'-r')=-0.35$ very similar to our
white dwarfs (see Fig.\ \ref{fig:ugr2}).  Figure \ref{fig:wd} shows the spectra
of two white dwarfs in our survey with the most unusually red $(u'-g')$ colors,
SDSS J074508.15+182630.0 (top) and SDSS J083303.03+365906.3 (bottom).  These
objects do not appear to be sdB subdwarfs because their spectra show only very
broad hydrogen Balmer lines.  It would be very interesting to know whether
these white dwarfs are unusually low mass white dwarfs, but detailed modeling
is beyond the scope of this paper.

	We search for proper motions in the USNOB1 and GSC2.3 catalogs, and
find proper motion measurements for 35 of the 44 white dwarfs, 20 of which are
significant at $>3\sigma$ level.  The average proper motion of the 20 white
dwarfs is 40 mas yr$^{-1}$ with an uncertainty of 7 mas yr$^{-1}$.  The late
B-type stars, by comparison, have no significant proper motion detections,
consistent with their inferred distances.  We calculate reduced proper motions
for the white dwarfs with proper motion measurements and find values ranging
$14<H_{g'}<18$ at $-0.7<(g'-i')_0<-0.5$, which places our objects in the main
body of white dwarfs observed by \citet{kleinman04} and \citet{kilic06}.


	Table \ref{tab:wd} lists the 44 spectroscopically identified white
dwarfs.  The columns include RA and Dec coordinates (J2000), $g'$ apparent
magnitude, $(u'-g')_0$ and $(g'-r')_0$ colors, and heliocentric radial
velocities $v_{helio}$.  We note that broad Balmer lines make for very poor
radial velocity determinations.  The objects are all DA white dwarfs with two
exceptions:  SDSS J111133.37+134639.8 is a DB white dwarf and SDSS
J151852.49+530121.8 is a DZ white dwarf with strong calcium H and K lines.

\section{UNUSUAL STARS IN DWARF GALAXIES}

\subsection{B Supergiants in the Leo A Dwarf}

	Leo A is an extremely metal-poor, gas-rich Im dwarf galaxy.  Stellar
population studies show that Leo A contains both very young and very old
stellar populations \citep{tolstoy98,schulte02,dolphin02,vansevicius04}.  To
date, stellar population studies of Leo A are based entirely on color-magnitude
diagrams, all of which reveal a striking ``blue plume'' of B giants possibly in
the galaxy.  Here, we discuss the first spectroscopic identifications of two B
giants definitely associated with the Leo A dwarf galaxy.  

	Two stars from our survey, SDSS J095915.12+304410.4 and SDSS
J095920.22+304352.7, match Leo A both in position and in velocity.  The stars
are located 1.2\arcmin\ and 2.0\arcmin, respectively, from the center of Leo A,
well within the 7\arcmin\ $\times$ 4.6\arcmin\ Holmberg diameter of the galaxy
\citep{mateo98}.  The stars have heliocentric radial velocity $+20\pm12$ and
$+32\pm12$ km s$^{-1}$, respectively, consistent at the 1$\sigma$ level with
the velocity of Leo A $+24\pm2$ km s$^{-1}$ measured from 21 cm observations
\citep{young96}.  The stars have apparent magnitude $g'=19.90\pm0.03$ and
$19.44\pm0.03$, respectively.  If the stars are physically associated with Leo
A, the galaxy's distance modulus $(m-M)_0 = 24.51 \pm 0.12$ \citep{dolphin02}
implies that the stars have luminosity $M_V \simeq -4.6$ and $-5.0$,
respectively.

\begin{figure}          
 \includegraphics[width=3.25in]{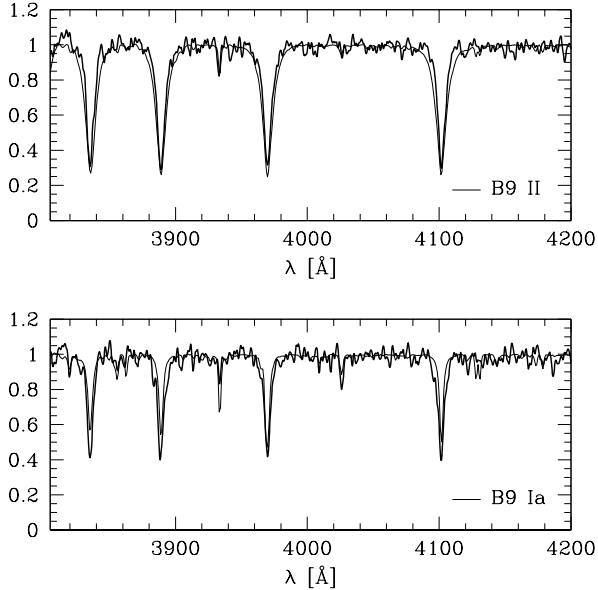}
 \figcaption{ \label{fig:b194}
        Spectra of SDSS J095915.12+304410.4 ({\it upper panel}) and SDSS 
J095920.22+304352.7 ({\it lower panel}) located in Leo A.  The observations are
convolved to match the 1.8 \AA\ resolution of the B9 II and B9 Ia MK standards 
\citep{gray03} overplotted as thin lines.}
 \end{figure}

	Interestingly, the spectra of the two stars in Leo A have unusually
narrow Balmer lines for stars in our sample; cross correlation with MK spectral
standards indicates that the stars are most likely luminosity class I or II B
supergiants.  Figure \ref{fig:b194} displays a portion of the spectra for SDSS
J095915.12+304410.4 ({\it upper panel}) and SDSS J095920.22+304352.7 ({\it
lower panel}), convolved to match the 1.8 \AA\ resolution of MK spectral
standards from \citet{gray03}.  The B9 II ($\gamma$ Lyr) and B9 Ia (HR1035) MK
standards are over-plotted as thin lines.  It is visually apparent that the
observed stars have Balmer line widths between those of the B II and B Ia
standards.  \citet{garrison84} give luminosities $M_V=-3.1$ for a B9 II star
and $M_V=-5.5$ for a B9 Ib star.  The luminosities we infer from the distance
to Leo A fall between these values, consistent with the spectra.  We conclude
the two stars are B supergiants in the blue plume of the Leo A dwarf galaxy.  
Such stars are $\sim$30 Myr old \citep{schaller92}, consistent with star
formation age estimates by others from color-magnitude diagrams. 

\subsection{UV-Bright BHB Star in the Draco Dwarf}

	By chance, another star from our survey is located in the Draco dwarf
galaxy.  The star, SDSS J172004.07+575110.8, has a spectral type of B9 and an
apparent magnitude of $g'=18.44\pm0.02$.  The star is also identified as
non-variable star \#517 in the classic \citet{baade61} paper.  The distance
modulus to Draco, $(m-M)_0 = 19.40 \pm 0.15$ \citep{bonanos04}, implies that
the star has $M_V \sim -1$, a more difficult luminosity to explain.  Unlike the
two stars in Leo A, the star in Draco has Balmer line widths inconsistent with
B giants.  We conclude that the most likely explanation for the star in Draco
is that it is a UV-bright, ``slow blue phase'' horizontal branch star.

	The position, velocity, and metallicity of the star in Draco match that
of the dwarf galaxy.  The star falls within 5\arcmin\ of the center of the
Draco, well within the 9\arcmin\ core radius of the galaxy \citep{irwin95}.  
The star's velocity $v_{RF} = -82 \pm 12$ km s$^{-1}$ is consistent at the
1$\sigma$ level with the velocity of Draco $-104 \pm21$ km s$^{-1}$
\citep{falco99}.  Finally, our estimate of the star's metallicity,
[Fe/H]$_{W_k}=-1.6\pm0.75$, is consistent with spectroscopic metallicity
measurements of Draco's stellar population that fall into two groups near
[Fe/H]$=-1.6\pm0.2$ and $-2.3\pm0.2$ \citep{shetrone01,lehnert92,kinman81,zinn78}.

	The star in Draco is probably not a main sequence B star, because there
is little evidence for young stars in color-magnitude diagrams of Draco
\citep[e.g.][]{bonanos04,klessen03,bellazzini02}.  It is possible that a B9
main sequence star in Draco is a blue straggler.  However, the luminosity of a
metal-poor B9 main sequence star is too low to place it at the distance of
Draco.  A main sequence star with $Z=0.001$ and $T_{eff}=10,500$ K has a mass
of 1.7 $M_\sun$ and an absolute magnitude $M_V(B9)=+1.6$ \citep{schaller92},
two-and-a-half magnitudes too faint to be at the distance of Draco.

	The star in Draco is also unlikely to be a normal BHB star.  The
horizontal branch of Draco is well observed and its stars are $20<V<21$
\citep{klessen03,bellazzini02}.  Moreover, a hot BHB star with spectral type B9
is an intrinsically faint star; the \citet{clewley04} $M_V(BHB)$ relation
yields $M_V(BHB)=+1.3$ which is two magnitudes too faint to be at the distance
Draco.

	Other possibilities, such as a blue-loop Cepheid or a post-AGB star,
are also unlikely.  Cepheids with masses $>5$ $M_{\sun}$ can travel out of the
instability strip on long blue loops \citep{bono00}, but massive stars are
unlikely to exist in Draco.  Post-AGB stars, stars in the process of blowing
off their outer layers to become white dwarfs, can have effective temperatures
of $10^4$ K but only for a short time.  Although there may be many more AGB
stars than BHB stars in Draco, the substantially shorter $10^3 - 10^4$ yr
timescale for a post-AGB star to have the correct effective temperature and
luminosity (P.\ Demarque, private communication) suggests that a longer-lived,
UV-bright star evolving off of the horizontal branch is a more plausible
explanation.

	The UV-bright phase is a slow-evolving, helium shell-burning phase that
occurs for BHB stars with small hydrogen envelopes.  Although the UV-bright
phase is more common in metal-rich stars, it occurs in metal-poor stars as
well.  \citet{yi97} stellar evolution tracks (see their Figure 1) show that
metal-poor BHB stars with $\sim0.05$ $M_\sun$ envelopes spend $10^7$ yrs at
effective temperatures around 10,000 K and $10^{2.4}$ $L_\sun$.  This model
provides the exact absolute magnitude $M_V(UV~BHB)=-1$ and spectral type needed
to place the star at the distance of Draco, and applies to stars with
metallicities ranging from [Fe/H]$=-1$ down to $-2.6$.  A recent study of BHB
stars in Draco identifies $\sim$50 BHB stars in the dwarf galaxy
\citep{klessen03}. If BHB stars spend 150 Myr on the BHB and 10 Myr in the
UV-bright phase, then we may expect a few BHB stars in the UV bright phase.

	If the star in Draco is a UV-bright BHB star, its spectrum should
indicate a low surface gravity.  We estimate the surface gravity of the star by
measuring the size and steepness of the Balmer jump \citep{kinman94}, and the
widths and the shapes of the Balmer lines \citep{clewley04}.  These independent
techniques indicate that the star is a low surface gravity star.  We conclude
that the star in Draco SDSS J172004.1+575111 is most likely UV-bright BHB star.

\section{CONCLUSIONS}

	In this paper we discuss our targeted survey for HVSs, a spectroscopic
survey of stars with late B-type colors that is now half complete.  Our survey
has discovered a total of four HVSs, or approximately 1-in-50 of our
candidates. The first two HVS discoveries are reported in \citet{brown06}.  
Here we report two new HVS discoveries:  HVS6 and possibly HVS7, traveling
with Galactic rest-frame velocities at least $+508\pm12$ and $+418\pm10$ km
s$^{-1}$, respectively.  Assuming the HVSs are main sequence B stars, they are
at distances $\sim$75 and $\sim$55 kpc, respectively, and have travel times
from the Galactic center consistent with their lifetimes.

	The remaining late B-type stars have metallicities and kinematics
consistent with a Galactic halo population of post main-sequence stars or blue
stragglers.  However, the line-of-sight velocity distribution shows a tail of
objects with large positive velocities.  This high velocity tail may be a mix
of low-velocity HVSs and high-velocity runaway stars; further theoretical and
observational work is needed to understand the nature of the high velocity
tail.

	Our survey includes many interesting objects besides HVSs.  
Approximately one-sixth of the objects are DA white dwarfs with unusually red
colors, possibly extremely low mass objects.

	Two of our objects are luminosity class I or II B supergiants in the
Leo A dwarf.  Our observations of these B supergiants provide the first
spectroscopic evidence for recent $\sim$30 Myr old star formation in Leo A.  
Another object is an unusual UV bright phase BHB star in the Draco dwarf.

	We are continuing our targeted HVS survey of late B-type stars in the
SDSS using the MMT telescope.  We are also using the Whipple 1.5m telescope to
obtain spectroscopy of brighter $15<g'<17$ late B-type objects.  Given our
current discovery rate, we expect to find perhaps another half dozen HVSs in
the coming months.  Follow-up high dispersion spectroscopy will provide precise
stellar parameters of these stars, and {\it Hubble Space Telescope}
observations will provide accurate proper motions.  Our goal is to discover
enough HVSs to allow us to place quantitative constraints on the stellar mass
function of HVSs, the origin of massive stars in the Galactic Center, and the
history of stellar interactions with the MBH.

\acknowledgements

	We thank R.\ Zinn and P.\ DeMarque for helpful conversations, and the
referee for a detailed report.  We thank M.\ Alegria, J.\ McAfee, and A.\
Milone for their assistance with observations obtained at the MMT Observatory,
a joint facility of the Smithsonian Institution and the University of Arizona.  
This project makes use of data products from the Sloan Digital Sky Survey,
which is managed by the Astrophysical Research Consortium for the Participating
Institutions.  This work was supported by W.\ Brown's Clay Fellowship and the
Smithsonian Institution.

{\it Facilities:} MMT (Blue Channel Spectrograph)


\begin{thebibliography}{78}
\expandafter\ifx\csname natexlab\endcsname\relax\def\natexlab#1{#1}\fi

\bibitem[{{Abt} {et~al.}(2002){Abt}, {Levato}, \& {Grosso}}]{abt02}
{Abt}, H.~A., {Levato}, H., \& {Grosso}, M. 2002, \apj, 573, 359

\bibitem[{{Adelman-McCarthy} {et~al.}(2006)}]{adelman06}
{Adelman-McCarthy}, J.~K. {et~al.} 2006, \apjs, 162, 38

\bibitem[{{Baade} \& {Swope}(1961)}]{baade61}
{Baade}, W. \& {Swope}, H.~H. 1961, \aj, 66, 300

\bibitem[{{Behr}(2003)}]{behr03}
{Behr}, B.~B. 2003, \apjs, 149, 67

\bibitem[{{Bellazzini} {et~al.}(2002){Bellazzini}, {Ferraro}, {Origlia},
  {Pancino}, {Monaco}, \& {Oliva}}]{bellazzini02}
{Bellazzini}, M., {Ferraro}, F.~R., {Origlia}, L., {Pancino}, E., {Monaco}, L.,
  \& {Oliva}, E. 2002, \aj, 124, 3222

\bibitem[{{Bellazzini} {et~al.}(2005){Bellazzini}, {Gennari}, \&
  {Ferraro}}]{bellazzini05}
{Bellazzini}, M., {Gennari}, N., \& {Ferraro}, F.~R. 2005, \mnras, 360, 185

\bibitem[{{Bellazzini} {et~al.}(2004){Bellazzini}, {Gennari}, {Ferraro}, \&
  {Sollima}}]{bellazzini04}
{Bellazzini}, M., {Gennari}, N., {Ferraro}, F.~R., \& {Sollima}, A. 2004,
  \mnras, 354, 708

\bibitem[{{Blaauw}(1961)}]{blaauw61}
{Blaauw}, A. 1961, \bain, 15, 265

\bibitem[{{Bonanos} {et~al.}(2004){Bonanos}, {Stanek}, {Szentgyorgyi},
  {Sasselov}, \& {Bakos}}]{bonanos04}
{Bonanos}, A.~Z., {Stanek}, K.~Z., {Szentgyorgyi}, A.~H., {Sasselov}, D.~D., \&
  {Bakos}, G.~{\'A}. 2004, \aj, 127, 861

\bibitem[{{Bono} {et~al.}(2000){Bono}, {Caputo}, {Cassisi}, {Marconi},
  {Piersanti}, \& {Tornamb{\`e}}}]{bono00}
{Bono}, G., {Caputo}, F., {Cassisi}, S., {Marconi}, M., {Piersanti}, L., \&
  {Tornamb{\`e}}, A. 2000, \apj, 543, 955

\bibitem[{{Brown} {et~al.}(2003){Brown}, {Allende Prieto}, {Beers}, {Wilhelm},
  {Geller}, {Kenyon}, \& {Kurtz}}]{brown03}
{Brown}, W.~R., {Allende Prieto}, C., {Beers}, T.~C., {Wilhelm}, R., {Geller},
  M.~J., {Kenyon}, S.~J., \& {Kurtz}, M.~J. 2003, \aj, 126, 1362

\bibitem[{{Brown} {et~al.}(2005{\natexlab{a}}){Brown}, {Geller}, {Kenyon}, \&
  {Kurtz}}]{brown05}
{Brown}, W.~R., {Geller}, M.~J., {Kenyon}, S.~J., \& {Kurtz}, M.~J.
  2005{\natexlab{a}}, \apjl, 622, L33

\bibitem[{{Brown} {et~al.}(2006){Brown}, {Geller}, {Kenyon}, \&
  {Kurtz}}]{brown06}
---. 2006, \apjl, 640, L35

\bibitem[{{Brown} {et~al.}(2005{\natexlab{b}}){Brown}, {Geller}, {Kenyon},
  {Kurtz}, {Allende Prieto}, {Beers}, \& {Wilhelm}}]{brown05b}
{Brown}, W.~R., {Geller}, M.~J., {Kenyon}, S.~J., {Kurtz}, M.~J., {Allende
  Prieto}, C., {Beers}, T.~C., \& {Wilhelm}, R. 2005{\natexlab{b}}, \aj, 130,
  1097

\bibitem[{{Callanan} {et~al.}(1998){Callanan}, {Garnavich}, \&
  {Koester}}]{callanan98}
{Callanan}, P.~J., {Garnavich}, P.~M., \& {Koester}, D. 1998, \mnras, 298, 207

\bibitem[{{Christodoulou} {et~al.}(1997){Christodoulou}, {Tohline}, \&
  {Keenan}}]{christodoulou97}
{Christodoulou}, D.~M., {Tohline}, J.~E., \& {Keenan}, F.~P. 1997, \apj, 486,
  810

\bibitem[{{Clewley} {et~al.}(2004){Clewley}, {Warren}, {Hewett}, {Norris}, \&
  {Evans}}]{clewley04}
{Clewley}, L., {Warren}, S.~J., {Hewett}, P.~C., {Norris}, J.~E., \& {Evans},
  N.~W. 2004, \mnras, 352, 285

\bibitem[{{Clewley} {et~al.}(2002){Clewley}, {Warren}, {Hewett}, {Norris},
  {Peterson}, \& {Evans}}]{clewley02}
{Clewley}, L., {Warren}, S.~J., {Hewett}, P.~C., {Norris}, J.~E., {Peterson},
  R.~C., \& {Evans}, N.~W. 2002, \mnras, 337, 87

\bibitem[{{Clewley} {et~al.}(2005){Clewley}, {Warren}, {Hewett}, {Norris},
  {Wilkinson}, \& {Evans}}]{clewley05}
{Clewley}, L., {Warren}, S.~J., {Hewett}, P.~C., {Norris}, J.~E., {Wilkinson},
  M.~I., \& {Evans}, N.~W. 2005, \mnras, 362, 349

\bibitem[{{Demarque} \& {Virani}(2006)}]{demarque06}
{Demarque}, P. \& {Virani}, S. 2006, preprint (astro-ph/0603326)

\bibitem[{{Dolphin} {et~al.}(2002){Dolphin}, {Saha}, {Claver}, {Skillman},
  {Cole}, {Gallagher}, {Tolstoy}, {Dohm-Palmer}, \& {Mateo}}]{dolphin02}
{Dolphin}, A.~E., {Saha}, A., {Claver}, J., {Skillman}, E.~D., {Cole}, A.~A.,
  {Gallagher}, J.~S., {Tolstoy}, E., {Dohm-Palmer}, R.~C., \& {Mateo}, M. 2002,
  \aj, 123, 3154

\bibitem[{{Dolphin} {et~al.}(2003){Dolphin}, {Saha}, {Skillman}, {Dohm-Palmer},
  {Tolstoy}, {Cole}, {Gallagher}, {Hoessel}, \& {Mateo}}]{dolphin03}
{Dolphin}, A.~E., {Saha}, A., {Skillman}, E.~D., {Dohm-Palmer}, R.~C.,
  {Tolstoy}, E., {Cole}, A.~A., {Gallagher}, J.~S., {Hoessel}, J.~G., \&
  {Mateo}, M. 2003, \aj, 125, 1261

\bibitem[{{Dray} {et~al.}(2005){Dray}, {Dale}, {Beer}, {Napiwotzki}, \&
  {King}}]{dray05}
{Dray}, L.~M., {Dale}, J.~E., {Beer}, M.~E., {Napiwotzki}, R., \& {King}, A.~R.
  2005, \mnras, 364, 59

\bibitem[{{Edelmann} {et~al.}(2005){Edelmann}, {Napiwotzki}, {Heber},
  {Christlieb}, \& {Reimers}}]{edelmann05}
{Edelmann}, H., {Napiwotzki}, R., {Heber}, U., {Christlieb}, N., \& {Reimers},
  D. 2005, \apjl, 634, L181

\bibitem[{{Eisenhauer} {et~al.}(2005)}]{eisenhauer05}
{Eisenhauer}, F. {et~al.} 2005, \apj, 628, 246

\bibitem[{{Falco} {et~al.}(1999){Falco}, {Kurtz}, {Geller}, {Huchra}, {Peters},
  {Berlind}, {Mink}, {Tokarz}, \& {Elwell}}]{falco99}
{Falco}, E.~E., {Kurtz}, M.~J., {Geller}, M.~J., {Huchra}, J.~P., {Peters}, J.,
  {Berlind}, P., {Mink}, D.~J., {Tokarz}, S.~P., \& {Elwell}, B. 1999, \pasp,
  111, 438

\bibitem[{{Fuentes} {et~al.}(2006){Fuentes}, {Stanek}, {Gaudi}, {McLeod},
  {Bogdanov}, {Hartman}, {Hickox}, \& {Holman}}]{fuentes06}
{Fuentes}, C.~I., {Stanek}, K.~Z., {Gaudi}, B.~S., {McLeod}, B.~A., {Bogdanov},
  S., {Hartman}, J.~D., {Hickox}, R.~C., \& {Holman}, M.~J. 2006, \apjl, 636,
  L37

\bibitem[{{Fukugita} {et~al.}(1996){Fukugita}, {Ichikawa}, {Gunn}, {Doi},
  {Shimasaku}, \& {Schneider}}]{fukugita96}
{Fukugita}, M., {Ichikawa}, T., {Gunn}, J.~E., {Doi}, M., {Shimasaku}, K., \&
  {Schneider}, D.~P. 1996, \aj, 111, 1748

\bibitem[{{Garrison}(1984)}]{garrison84}
{Garrison}, R.~F., ed. 1984, {The MK process and stellar classification}
  (Toronto: DDO), 275

\bibitem[{{Ginsburg} \& {Loeb}(2006)}]{ginsburg06}
{Ginsburg}, I. \& {Loeb}, A. 2006, preprint (astro-ph/0510574)

\bibitem[{{Gnedin} {et~al.}(2005){Gnedin}, {Gould}, {Miralda-Escud{\'e}}, \&
  {Zentner}}]{gnedin05}
{Gnedin}, O.~Y., {Gould}, A., {Miralda-Escud{\'e}}, J., \& {Zentner}, A.~R.
  2005, \apj, 634, 344

\bibitem[{{Gray} {et~al.}(2003){Gray}, {Corbally}, {Garrison}, {McFadden}, \&
  {Robinson}}]{gray03}
{Gray}, R.~O., {Corbally}, C.~J., {Garrison}, R.~F., {McFadden}, M.~T., \&
  {Robinson}, P.~E. 2003, \aj, 126, 2048

\bibitem[{{Gualandris} {et~al.}(2004){Gualandris}, {Portegies Zwart}, \&
  {Eggleton}}]{gualandris04}
{Gualandris}, A., {Portegies Zwart}, S., \& {Eggleton}, P.~P. 2004, \mnras,
  350, 615

\bibitem[{{Gualandris} {et~al.}(2005){Gualandris}, {Zwart}, \&
  {Sipior}}]{gualandris05}
{Gualandris}, A., {Zwart}, S.~P., \& {Sipior}, M.~S. 2005, \mnras, 363, 223

\bibitem[{{Harris} {et~al.}(2003)}]{harris03}
{Harris}, H.~C. {et~al.} 2003, \aj, 126, 1023

\bibitem[{{Heber} {et~al.}(2003){Heber}, {Edelmann}, {Lisker}, \&
  {Napiwotzki}}]{heber03}
{Heber}, U., {Edelmann}, H., {Lisker}, T., \& {Napiwotzki}, R. 2003, \aap, 411,
  L477

\bibitem[{{Hills}(1988)}]{hills88}
{Hills}, J.~G. 1988, \nat, 331, 687

\bibitem[{{Hills}(1991)}]{hills91}
---. 1991, \aj, 102, 704

\bibitem[{{Hirsch} {et~al.}(2005){Hirsch}, {Heber}, {O'Toole}, \&
  {Bresolin}}]{hirsch05}
{Hirsch}, H.~A., {Heber}, U., {O'Toole}, S.~J., \& {Bresolin}, F. 2005, \aap,
  444, L61

\bibitem[{{Hogg} {et~al.}(2005){Hogg}, {Blanton}, {Roweis}, \&
  {Johnston}}]{hogg05}
{Hogg}, D.~W., {Blanton}, M.~R., {Roweis}, S.~T., \& {Johnston}, K.~V. 2005,
  \apj, 629, 268

\bibitem[{{Holley-Bockelmann} {et~al.}(2006){Holley-Bockelmann}, {Sigurdsson},
  {Mihos}, {Feldmeier}, {Ciardullo}, \& {McBride}}]{holley06}
{Holley-Bockelmann}, K., {Sigurdsson}, S., {Mihos}, C.~J., {Feldmeier}, J.~J.,
  {Ciardullo}, R., \& {McBride}, C. 2006, preprint (astro-ph/0512344)

\bibitem[{{Irwin} \& {Hatzidimitriou}(1995)}]{irwin95}
{Irwin}, M. \& {Hatzidimitriou}, D. 1995, \mnras, 277, 1354

\bibitem[{{Kilic} {et~al.}(2006)}]{kilic06}
{Kilic}, M. {et~al.} 2006, \aj, 131, 582

\bibitem[{{Kinman} {et~al.}(1981){Kinman}, {Kraft}, \& {Suntzeff}}]{kinman81}
{Kinman}, T.~D., {Kraft}, R.~P., \& {Suntzeff}, N.~B. 1981, in ASSL Vol. 88:
  Physical Processes in Red Giants, 71--76

\bibitem[{{Kinman} {et~al.}(1994){Kinman}, {Suntzeff}, \& {Kraft}}]{kinman94}
{Kinman}, T.~D., {Suntzeff}, N.~B., \& {Kraft}, R.~P. 1994, \aj, 108, 1722

\bibitem[{{Kleinman} {et~al.}(2004)}]{kleinman04}
{Kleinman}, S.~J. {et~al.} 2004, \apj, 607, 426

\bibitem[{{Klessen} {et~al.}(2003){Klessen}, {Grebel}, \&
  {Harbeck}}]{klessen03}
{Klessen}, R.~S., {Grebel}, E.~K., \& {Harbeck}, D. 2003, \apj, 589, 798

\bibitem[{{Kurtz} \& {Mink}(1998)}]{kurtz98}
{Kurtz}, M.~J. \& {Mink}, D.~J. 1998, \pasp, 110, 934

\bibitem[{{Lehnert} {et~al.}(1992){Lehnert}, {Bell}, {Hesser}, \&
  {Oke}}]{lehnert92}
{Lehnert}, M.~D., {Bell}, R.~A., {Hesser}, J.~E., \& {Oke}, J.~B. 1992, \apj,
  395, 466

\bibitem[{{Leonard}(1991)}]{leonard91}
{Leonard}, P.~J.~T. 1991, \aj, 101, 562

\bibitem[{{Leonard}(1993)}]{leonard93}
{Leonard}, P.~J.~T. 1993, in ASP Conf.\ Ser.\ 45, Luminous High-Latitude Stars,
  ed. D.~D. Sasselov, 360

\bibitem[{{Levin}(2005)}]{levin05}
{Levin}, Y. 2005, preprint (astro-ph/0508193)

\bibitem[{{Liebert} {et~al.}(2004){Liebert}, {Bergeron}, {Eisenstein},
  {Harris}, {Kleinman}, {Nitta}, \& {Krzesinski}}]{liebert04}
{Liebert}, J., {Bergeron}, P., {Eisenstein}, D., {Harris}, H.~C., {Kleinman},
  S.~J., {Nitta}, A., \& {Krzesinski}, J. 2004, \apjl, 606, L147

\bibitem[{{Lynn} {et~al.}(2004){Lynn}, {Keenan}, {Dufton}, {Saffer},
  {Rolleston}, \& {Smoker}}]{lynn04}
{Lynn}, B.~B., {Keenan}, F.~P., {Dufton}, P.~L., {Saffer}, R.~A., {Rolleston},
  W.~R.~J., \& {Smoker}, J.~V. 2004, \mnras, 349, 821

\bibitem[{{Martin}(2004)}]{martin04}
{Martin}, J.~C. 2004, \aj, 128, 2474

\bibitem[{{Martin}(2006)}]{martin06}
---. 2006, preprint (astro-ph/0603238)

\bibitem[{{Mateo}(1998)}]{mateo98}
{Mateo}, M.~L. 1998, \araa, 36, 435

\bibitem[{{Monet} {et~al.}(2003)}]{monet03}
{Monet}, D.~G. {et~al.} 2003, \aj, 125, 984

\bibitem[{{O'Connell}(1973)}]{oconnell73}
{O'Connell}, W.~O.~R. 1973, \aj, 78, 1074

\bibitem[{{O'Toole} {et~al.}(2006){O'Toole}, {Napiwotzki}, {Heber}, {Drechsel},
  {Frandsen}, {Grundahl}, \& {Bruntt}}]{otoole06}
{O'Toole}, S.~J., {Napiwotzki}, R., {Heber}, U., {Drechsel}, H., {Frandsen},
  S., {Grundahl}, F., \& {Bruntt}, H. 2006, Baltic Astronomy, 15, 61

\bibitem[{{Peterson} {et~al.}(1995){Peterson}, {Rood}, \&
  {Crocker}}]{peterson95}
{Peterson}, R.~C., {Rood}, R.~T., \& {Crocker}, D.~A. 1995, \apj, 453, 214

\bibitem[{{Portegies Zwart}(2000)}]{portegies00}
{Portegies Zwart}, S.~F. 2000, \apj, 544, 437

\bibitem[{{Poveda} {et~al.}(1967){Poveda}, {Ruiz}, \& {Allen}}]{poveda67}
{Poveda}, A., {Ruiz}, J., \& {Allen}, C. 1967, Bol.\ Obs\ Tonantzintla
  Tacubaya, 4, 860

\bibitem[{{Schaller} {et~al.}(1992){Schaller}, {Schaerer}, {Meynet}, \&
  {Maeder}}]{schaller92}
{Schaller}, G., {Schaerer}, D., {Meynet}, G., \& {Maeder}, A. 1992, \aaps, 96,
  269

\bibitem[{{Schlegel} {et~al.}(1998){Schlegel}, {Finkbeiner}, \&
  {Davis}}]{schlegel98}
{Schlegel}, D.~J., {Finkbeiner}, D.~P., \& {Davis}, M. 1998, \apj, 500, 525

\bibitem[{{Schulte-Ladbeck} {et~al.}(2002){Schulte-Ladbeck}, {Hopp},
  {Drozdovsky}, {Greggio}, \& {Crone}}]{schulte02}
{Schulte-Ladbeck}, R.~E., {Hopp}, U., {Drozdovsky}, I.~O., {Greggio}, L., \&
  {Crone}, M.~M. 2002, \aj, 124, 896

\bibitem[{{Shetrone} {et~al.}(2001){Shetrone}, {C{\^o}t{\'e}}, \&
  {Sargent}}]{shetrone01}
{Shetrone}, M.~D., {C{\^o}t{\'e}}, P., \& {Sargent}, W.~L.~W. 2001, \apj, 548,
  592

\bibitem[{{Tolstoy} {et~al.}(1998)}]{tolstoy98}
{Tolstoy}, E. {et~al.} 1998, \aj, 116, 1244

\bibitem[{{van Woerden}(1993)}]{vanwoerden93}
{van Woerden}, H. 1993, in ASP Conf. Ser. 45: Luminous High-Latitude Stars, ed.
  D.~D. {Sasselov}, 11

\bibitem[{{Vansevi{\v c}ius} {et~al.}(2004)}]{vansevicius04}
{Vansevi{\v c}ius}, V. {et~al.} 2004, \apjl, 611, L93

\bibitem[{{Wilhelm} {et~al.}(1999){Wilhelm}, {Beers}, \& {Gray}}]{wilhelm99a}
{Wilhelm}, R., {Beers}, T.~C., \& {Gray}, R.~O. 1999, \aj, 117, 2308

\bibitem[{{Wilkinson} \& {Evans}(1999)}]{wilkinson99}
{Wilkinson}, M.~I. \& {Evans}, N.~W. 1999, \mnras, 310, 645

\bibitem[{{Worthey} {et~al.}(1994){Worthey}, {Faber}, {Gonzalez}, \&
  {Burstein}}]{worthey94}
{Worthey}, G., {Faber}, S.~M., {Gonzalez}, J.~J., \& {Burstein}, D. 1994,
  \apjs, 94, 687

\bibitem[{{Yanny} {et~al.}(2000)}]{yanny00}
{Yanny}, B. {et~al.} 2000, \apj, 540, 825

\bibitem[{{Yi}, {Demarque}, \& {Kim}(1997)}]{yi97}
{Yi}, S., {Demarque}, P., \& {Kim}, Y.-C. 1997, \apj, 482, 677

\bibitem[{{Young} \& {Lo}(1996)}]{young96}
{Young}, L.~M. \& {Lo}, K.~Y. 1996, \apj, 462, 203

\bibitem[{{Yu} \& {Tremaine}(2003)}]{yu03}
{Yu}, Q. \& {Tremaine}, S. 2003, \apj, 599, 1129

\bibitem[{{Zinn}(1978)}]{zinn78}
{Zinn}, R. 1978, \apj, 225, 790

\end{thebibliography}


\clearpage

\begin{deluxetable}{lccccccl}           
\tablewidth{0pt}
\tablecaption{HYPERVELOCITY STARS\label{tab:hvs}}
\tablecolumns{8}
\tablehead{
  \colhead{ID} & \colhead{$l$} & \colhead{$b$} & \colhead{$g'$} &
  \colhead{$v_{RF}$} & \colhead{$d$} & \colhead{$t_{GC}$} & \colhead{Catalog} \\
  \colhead{} & \colhead{{\small deg}} & \colhead{{\small deg}} &
  \colhead{{\small mag}} & \colhead{{\small km s$^{-1}$}} &
  \colhead{{\small kpc}} & \colhead{{\small Myr}} & \colhead{}
}
	\startdata
HVS1 & 227.3 & ~31.3 & 19.8 & +709 & 110 & 160 & SDSS J090745.0+024507$^1$ \\
HVS2 & 176.0 & ~47.1 & 18.8 & +717 & ~19 & ~32 & US 708$^2$ \\
HVS3 & 263.0 & -40.9 & 16.2 & +548 & ~61 & 100 & HE 0437-5439$^3$ \\
HVS4 & 194.8 & ~42.6 & 18.4 & +563 & ~75 & 130 & SDSS J091301.0+305120$^4$ \\
HVS5 & 146.3 & ~38.7 & 17.9 & +643 & ~55 & ~90 & SDSS J091759.5+672238$^4$ \\
HVS6 & 243.1 & ~59.6 & 19.1 & +508 & ~75 & 160 & SDSS J110557.45+093439.5 \\
HVS7\tablenotemark{*} & 263.8 & ~57.9 & 17.7 & +418 & ~55 & 120 & SDSS 
J113312.12+010824.9 \\
        \enddata
\tablenotetext{*}{Probable HVS.}
\tablecomments{HVS4 - HVS7 are from this targeted HVS survey.}
\tablerefs{ (1) \citet{brown05}; (2) \citet{hirsch05}; (3) \citet{edelmann05}; 
(4) \citet{brown06} }
 \end{deluxetable}

\begin{deluxetable}{cccccrr}		
\tabletypesize{\small}
\tablewidth{0pt}
\tablecaption{LATE B OBJECTS\label{tab:bhb}}
\tablecolumns{7}
\tablehead{
	\colhead{RA} & \colhead{Dec} &
	\colhead{$g'$} & \colhead{$(u'-g')_0$} & \colhead{$(g'-r')_0$} &
	\colhead{$v_{helio}$} & \colhead{[Fe/H]$_{W_k}$} \\
	\colhead{J2000} & \colhead{J2000} &
	\colhead{mag} & \colhead{mag} & \colhead{mag} &
	\colhead{km s$^{-1}$} & \colhead{}
}
	\startdata
 0:02:33.82 &  -9:57:06.8 & $18.578\pm0.021$ & $0.753\pm0.040$ & $-0.328\pm0.040$ & $ -88\pm10$ & $ 0.0\pm0.9$ \\
 0:05:28.14 & -11:00:10.1 & $19.271\pm0.042$ & $1.007\pm0.081$ & $-0.275\pm0.047$ & $-123\pm12$ & $-1.8\pm1.0$ \\
 0:07:52.01 &  -9:19:54.3 & $17.440\pm0.017$ & $1.016\pm0.036$ & $-0.276\pm0.039$ & $-119\pm11$ & $-1.6\pm0.6$ \\
	\enddata 
  \tablecomments{Table \ref{tab:bhb} is presented in its entirety in the
electronic edition of the Astrophysical Journal.  A portion is shown here for
guidance and content.}
 \end{deluxetable}

\begin{deluxetable}{cccccr}		
\tablewidth{0pt}
\tablecaption{WHITE DWARFS\label{tab:wd}}
\tablecolumns{6}
\tablehead{
	\colhead{RA} & \colhead{Dec} & \colhead{$g'$} & 
	\colhead{$(u'-g')_0$} & \colhead{$(g'-r')_0$} & \colhead{$v_{helio}$} \\
	\colhead{J2000} & \colhead{J2000} & \colhead{mag} &
	\colhead{mag} & \colhead{mag} & \colhead{km s$^{-1}$}
}
	\startdata
 0:28:03.34 &  -0:12:13.4 & $18.414\pm0.019$ & $0.517\pm0.032$ & $-0.418\pm0.025$ & $  97\pm43$ \\
 1:00:44.69 &  -0:50:34.1 & $20.111\pm0.062$ & $0.577\pm0.097$ & $-0.300\pm0.068$ & $  73\pm47$ \\
 1:06:57.83 & -10:08:39.3 & $19.417\pm0.025$ & $0.525\pm0.072$ & $-0.366\pm0.035$ & $ -23\pm47$ \\
	\enddata 
  \tablecomments{Table \ref{tab:wd} is presented in its entirety in the
electronic edition of the Astrophysical Journal.  A portion is shown here for
guidance and content.}
 \end{deluxetable}

\end{document}